# Microscopic Theory of Entropic Bonding for Colloidal Crystal Prediction

Thi Vo[1] and Sharon C. Glotzer,[1,2]*

[1] Department of Chemical Engineering, University of Michigan – Ann Arbor

[2] Biointerfaces Institute, University of Michigan – Ann Arbor

*To whom correspondence should be addressed; E-mail: sglotzer@umich.edu.

Entropy alone can self-assemble hard particles into colloidal crystals of remarkable complexity whose structures are the same as atomic and molecular crystals, but with larger lattice spacings. Although particle-based molecular simulation is a powerful tool for predicting self-assembly by exploring phase space, it is not yet possible to predict colloidal crystal structures *a priori* from particle shape as we can for atomic crystals based on electronic valency. Here we present such a first principles theory. By directly calculating and minimizing excluded volume within the framework of statistical mechanics, we describe the directional forces that emerge between hard shapes in familiar terms used to describe chemical bonds. Our theory predicts thermodynamically preferred structures for four families of hard polyhedra that match, in every instance, previous simulation results. The success of this first principles approach to entropic colloidal crystal structure prediction furthers fundamental understanding of both entropically driven crystallization and conceptual pictures of bonding in matter.

**Introduction**

In his 1704 treatise, *Opticks*, Sir Isaac Newton wrote of an "attractive" force that holds particles of matter together. By observing that matter stays together in a host of different everyday situations, he inferred the presence of what we now refer to as chemical bonds, one of the most fundamental concepts in science. Chemical bonds originate in electronic interactions. Together with thermodynamics, chemical bonds allow us to predict and understand materials structure and properties, including the way in which atoms arrange to form crystals, from the ionically bonded simple lattice of table salt to the covalently bonded structure of diamond to multicomponent alloys with many thousands of atoms in a unit cell held together by delocalized metallic bonding.

Like atoms, nanoparticles can self-assemble under thermodynamic driving forces into crystal structures of extraordinary complexity and diversity (1–6). What "force" would Newton infer from observing these "colloidal" crystals? From his writings, he would undoubtedly infer an attraction – a bond – between particles as for atoms, but of a different origin and at larger length, time, and energy scales. Indeed, DNA linkers, ligand-ligand pairs, protein-antibody pairs, and opposite charges are all examples of bonding elements exploited to self-assemble colloidal crystals of nanoparticles, which effectively serve as "atoms." But what would Newton make of colloidal crystals self-assembled from hard particles whose only attribute, like billiard balls, is the inability to overlap?

Remarkably, many reported hard particle colloidal crystals are isostructural to known atomic crystals with covalent, ionic and metallic bonding (7–13). For example, all 14 Bravais lattice types (7), Frank-Kasper (7), clathrate structures (11), open lattices (14), multiple quasicrystals (7, 8), and crystals with as many as 432 particles in the unit cell (11) self-assemble from disordered fluid phases in computer simulations of hard, otherwise non-interacting polyhedra. Experiments likewise find colloidal crystals of hard particles, though the diversity of structures reported to date has been limited by current challenges in making diverse shapes devoid of any energetic interactions (12, 15, 16). Hard particle crystals – in where there is no potential energy – are stabilized by entropy maximization, which, counterintuitively, is achieved when the ordered crystal has more available microstates than the disordered fluid (17–21). In hard particle systems, no explicit bonding elements are required for crystallization. Instead, entropy maximization leads to the emergence of effective, attractive entropic forces with directionality that arises from particle shape and crowding. This directionality creates local valence and, ultimately, long-range order (22,23). The extent to which the same complexity and diversity of structure is possible from chemical bonding and from entropic "bonding" is unexpected and demonstrates that emergent phenomena such as crystal self-assembly are agnostic to the origin of these bonds, provided the system is subject to the laws of thermodynamics. We assert that if entropic bonds are to be considered a useful construct like chemical bonds, then crystal structure prediction using *ab initio* methods should be possible, in the same way that atomic crystal structures can be predicted from a knowledge of chemical bonding to solve Schrodinger's Equation (24,25).

Here, we report an *ab initio* theory of entropic bonding that does not involve computer simulations but instead predicts thermodynamically preferred crystal structures directly from particle shape and a mean-field description of directional entropic forces. Using traditional statistical mechanics, we develop a framework for calculating entropic bonds between arbitrarily shaped particles. We test our theory on several challenging examples to show that our method correctly selects the thermodynamically preferred crystal structures reported in simulations. For brevity, we limit our discussion to single-component crystals composed of hard polyhedral particles, although generalization to multi-component crystals of arbitrarily shaped hard or patchy particles (26, 27) is foreseeable. Lastly, by comparing similarities in mathematical form for our statistical mechanical theory with quantum theory, we posit that theoretical ideas from quantum mechanics might be useful, if only by analogy, for understanding colloidal systems and *vice versa*. The success of our theory not only provides a new way to predict colloidal self-assembly, but also presents a potentially unifying framework that bridges our understanding of crystallization across different length scales.

**Results**

Our theory builds upon two fundamentals tenets of statistical mechanics. First, thermodynamic systems evolve to the most probable state. Second, the probability of any microstate – in this case, a unique arrangement of particles – depends on the energy of that microstate. Consider a system of $N$ hard particles of identical size and shape contained within a fixed volume $V$. Because hard particles don't interact, there is no energy and thus all microstates are equally probable. The most probable thermodynamic state is then the one with the largest number ($\Omega$) of microstates, or, equivalently, the maximum entropy $S$ via Boltzmann's expression $S = k_B \ln \Omega$.

For hard particle systems, maximizing entropy is equivalent to maximizing all of the empty (so-called "free") volume into which a new particle can be placed without overlapping another particle (20,21). Maximizing free volume is, in turn, achieved by minimizing the *excluded* volume around each particle (17,20). In lieu of directly calculating excluded volume (and thus entropy) in our system, we approximate excluded volume using a mathematical *ansatz* described below. In this way, we obtain an eigenvalue equation whose solution gives the free energy of a colloidal crystal structure without need for thermodynamic integration (34,35), a routine method for free energy calculation that relies on simulation. In what follows, we present the key results in our derivation, and leave the mathematical details to the Supporting Materials (SM).

**Pseudoparticle *Ansatz***

We propose an *ansatz* in which all empty space within the system of *N* hard anisotropic particles in a volume *V* is filled with $N_{pP}$ fictitious pseudoparticles (pP), each of which is much smaller than a particle, can freely overlap each other, and is attracted to anisotropic particles with an interaction energy we derive below. To avoid confusion, we hereafter refer to the anisotropic particles as nanoparticles (NPs), which are not to be confused with the fictitious pPs. In the limit of low NP density, the pP spatial distribution will be roughly uniform. At intermediate and high NP densities, the pP density distribution will become spatially dependent, conforming to the NP shape as the excluded volume becomes more and more spatially heterogeneous. Testing different NP arrangements and selecting for those that optimize the pP density distribution will yield the colloidal crystal structure with the maximum entropy among the set of structures tested. By representing the effective entropic forces by explicit forces between NPs encapsulated in a mean-field pP-NP interaction potential, we will predict the thermodynamically preferred structure that would otherwise necessitate computer experiments using Monte Carlo or molecular dynamics simulation.

**Interaction Between Nanoparticles and Fictitious Pseudoparticles**

We perform a self-consistent mean-field approximation (28) in the $[N, N_{pP}, V, T]$ statistical mechanical ensemble where we integrate over all NPs except for a single reference NP (or pP) to obtain the following pP-NP interaction potential (see Section II of the SM for details),

$$\beta U(r) = [\ln \rho_{pP}(r) - \beta \mu_{pP}]r^{-2} + \sum_{i}^{N} U_{core}(r) \qquad (1)$$

where $\beta = 1/k_B T$, $\rho_{pP}(r)$ is the spatially dependent, mean-field density distribution of the pPs, $\mu_{pP}$ is the chemical potential of the pPs, $U_{core}$ represents a hard-core repulsion preventing pPs from overlapping a NP, and *r* is the distance between a pP and a NP. Eq. 1 describes a pairwise interaction *U(r)* with potential wells (indicating attraction between NPs and pPs) that deepen with increasing $\mu_{pP}$ (Fig. 1c). A small $\mu_{pP}$ implies that it is easy to place more pPs into the system, and is expected in the dilute NP limit, where entropic forces are weak and largely isotropic. However, $\mu_{pP}$ increases with increasing density of NPs as it becomes harder to find a location to accommodate a new pP. As a result, the pP density distribution develops a strong spatial dependence. Optimizing this spatially dependent pP density distribution – i.e., maximizing the total pP-NP interaction (free) energy -- yields the most probable thermodynamic state of the system. This optimization maximizes the entropy of the system of NPs.

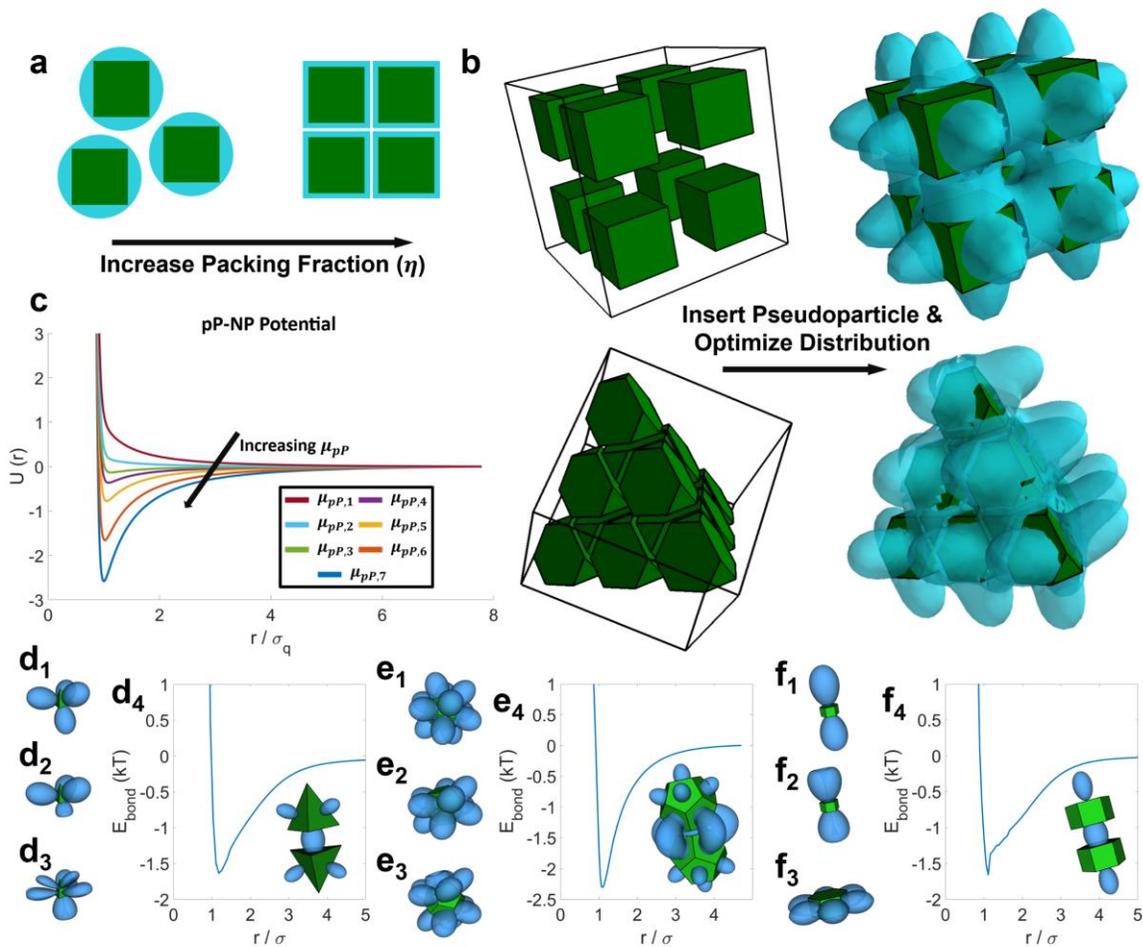

Figure 1: a) - c) Schematic of theoretical framework. a) Change in pP distribution about a shape as a function of packing fraction. Dilute (left): shape freely rotates and pPs occupy a volume equivalent to the shape's circumsphere size (drawn in blue). High packing fraction (right): shape crowding reduces the accessible space and pPs occupy a conformal region relative to particle shape. b) Visual representation of entropic bonding orbitals (high local pP probability density) calculated in systems of polyhedra. Crowding of shapes by bounding box (left) is mathematically represented by NPs "glued" together by pPs (right). c) Functional form of pP-NP potential (Eq. 1) derived from mean-field approximation. Shape orbitals and molecular orbitals for d) tetrahedron, e) dodecahedron, and f) hexagonal prism. Subplots 1-3 for d)-f) show the first three single particle shape orbitals for each shape. Subplot 4 for d)-f) shows the equilibrium (free) energy as a function of center-to-center separation distance at the predicted most stable orientation of a pair of polyhedra. Insets show the predicted lowest (free) energy (highest entropy) configuration (both position and orientation). All polyhedra prefer face-face alignment with equilibrium bond lengths of $\sim 1.1\sigma$. Tetrahedron and hexagonal prisms exhibit bond (free) energies of $\sim 1.5 – 1.6\ kT$. For the pair of dodecahedra, secondary stabilization (from local facets near the main face-face contact) lowers the bond (free) energy to $\sim 2.3\ kT$. The LCSO used in solving for the entropic bonding orbitals employs the lowest shape orbital for each polyhedron except for the dodecahedron.

It is worthwhile noting here that Eq. 1 results from a rearrangement of probability distribution functions (Eq. II.8 and II.13 in SM) obtained from our mean-field treatment. We derive analogous equations (Eq. II.9 and II.12 in the SM) for a lattice model by explicitly maximizing the entropy in a case where excluded volume can be calculated directly without the pP *ansatz*.

**From Mean-Field Interaction Potential to Eigenvalue Equation**

Eq. 1 describes how the pseudoparticles distribute themselves around a single NP. We next apply this equation to a system of $N$ NPs to calculate the optimal pP distribution for a given arrangement of NPs (e.g. colloidal crystal structure). To do that, we perform a material balance for a system of $N$ NPs and $N_{pP}$ pPs. The material balance can be constructed using the convection-diffusion equation (Eq. III.1 in SM), commonly employed in transport phenomena. Assuming steady state, with $U$ as in Eq. 1, gives

$$[\nabla^2 + \beta\nabla^2 U - (\beta\nabla U)^2]\rho_{pP} = E\rho_{pP} \qquad (2)$$

Eq. 2 is an eigenvalue equation for the spatially dependent pseudoparticle density distribution. The transformation operator on the left side of Eq. 2 contains a diffusive term, and (free) energy terms arising from the pP-NP interactions. Consequently, the eigenvalues $E$ obtained from solving this equation can be interpreted as free energies for a given arrangement of NPs, such as a crystal structure. Comparison of the computed eigenvalues for NPs arranged on different lattices predicts the crystal structure with the lowest total (free) energy of all tested crystal structures; by construction, this structure will have the highest entropy (Fig. 1d-f, 2 - 4). Solving this eigenvalue equation is non-trivial, however, and requires approximations and a series expansion of the local pP density in a coordinate system with symmetries consistent with those of the NP shape, as shown below.

**Shape Orbitals**

Because Eq. 2 is an eigenvalue equation, we can solve it using numerical methods found in quantum mechanical (QM) codes that solve the Schrodinger equation and calculate energies of atomic crystal structures. The first step involves expressing the pP probability density functional as a series of "shape harmonics" that explicitly accounts for the particle's shape, similar to the spherical harmonics used in QM (see details in Section IV of SM). The resulting shape harmonics are functions of two angular components. We call these shape harmonics "shape orbitals," in analogy with the correspondence between spherical harmonics and atomic orbitals. Fig. $1d_{1-3}$ - $f_{1-3}$ shows the first three shape orbitals for a tetrahedron, dodecahedron, and hexagonal prism. We take a linear combination of shape orbitals ("LCSO") as a trial solution to solve for the optimal configurations of pairs of neighboring polyhedra (25). The solution entails optimizing orbital overlap for a given lattice, considering all possible relative orientations of all NPs. Details of the calculation are given in Section V of the SM. For all shapes considered here except the dodecahedron, trial solutions using only the lowest order shape orbitals were sufficient to obtain accurate results. For the dodecahedron, the next lowest order shape orbital was required.

**Calculating the Entropic Bond for Polyhedral Nanoparticles**

For a pair of polyhedral NPs, solving Eq. 2 predicts the most stable configuration in a crowded system of NPs to be that which favors face-face alignment between polyhedra, in agreement with previous simulations (10,22,23,29). Strong facet alignment is indicated by high local density of pPs, and results from minimization of the total pP-NP free energy. Additionally, we find distinct minima when plotting free energy as a function of center-to-center separation between NPs (Fig. $1d_4$ - $f_4$). This preference for both a specific relative orientation and separation between NPs arises from optimizing shape-orbital overlaps between neighboring particles, and is analogous to the bond angles and bond lengths between atoms arising from optimizing electronic orbital overlaps (25,30). For the packing fractions considered, the "bond strengths" for face-aligned polyhedra lie between $1.5 - 1.6\ kT$, with a corresponding "bond length" of $1.1\sigma$, where $\sigma$ is the in-sphere diameter of each polyhedron. Stronger bonds result at higher densities. Dodecahedra exhibit a slightly stronger bond of $\sim 2.3\ kT$ as a result of additional orbital overlaps between small facets (Fig. 1e$_4$, inset), suggesting a potential design strategy for stabilizing specific surface contacts (face-to-face alignment) by introducing additional smaller facets near the primary, desired contact face. The calculations employed in Fig. 1 provide a quantitative, microscopic description of the entropic bond in terms of a local high density of pseudoparticles, and are readily extensible to particles of different shapes and sizes.

**Crystal Structure Prediction**

To illustrate the application of the entropic bond theory to colloidal crystal prediction, we begin with a simple example and compute the optimal arrangement of hard cubes at packing fraction $\eta = 0.55$ in three simple lattices – body-centered cubic (BCC), face-centered cubic (FCC) and simple cubic (SC). Fig. 2a shows the first three shape orbitals and bonding energy for cubes, employing a similar calculation performed in Fig. 1d-f. Fig. 2b shows the computed lattice free energy of formation, indicating that cubes are more stable within the SC lattice as compared to both BCC and FCC, as expected (9, 31). Visualization of the bonding orbitals (Fig. 2c-e) reveals why. We see that cubes in an FCC arrangement suffer a deficit of two face-face bonding orbitals per cube. While the FCC lattice gains edge-edge bonding between neighboring cubes, the gain is not sufficient to offset the loss of face-face bonding relative to the SC arrangement. Similarly, the center cube in the BCC lattice gains 12 edge-edge bonding contacts but loses all six of its face-face bonds. Since edge-edge bonding is lower in entropy than face-face bonding, BCC is significantly less favored for cubes relative to both SC and FCC (Fig. 2b).

Entropic bond theory can predict solid-solid transitions as particle shape is modified. Consider the truncated tetrahedron shape family (29). Simulation studies reported the self-assembly of a series of crystal structures, shown in Fig. 3a, as the corners of a regular tetrahedron are truncated (full truncation produces a regular octahedron) (29). Fig. 3b-c shows a direct comparison between the lowest free energy crystal structure predicted using our framework (colors in Fig. 3b match the corresponding unit cells in Fig. 3a) with the transitions reported in Damasceno *et al.* (29) shown in black lines, indicating excellent agreement between theory and simulation.

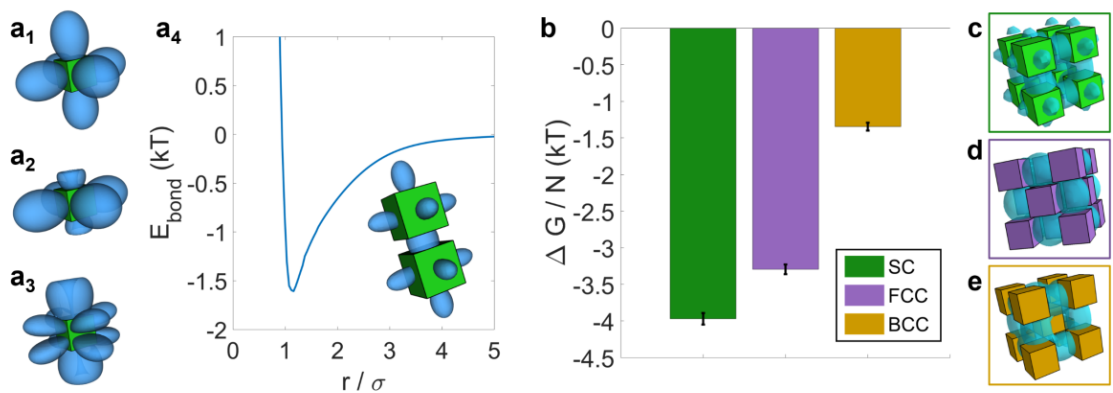

Figure 2: Shape orbitals and a bonding orbital for cubes. $a_1$) - $a_3$) show the first three single-particle shape orbitals for the cube. $a_4$) shows the equilibrium (free) energy as a function of center-to-center separation distance at the predicted most stable orientation of a pair of polyhedra. Inset shows the predicted lowest energy configuration (both position and orientation). Phase prediction for cubes. b) Free energy per cube within each respective lattice. Visualization of cubes with bonding orbital in c) simple cubic (SC). d) face-centered cubic (FCC) and e) body-centered cubic (BCC). The LCSO approximation employs the first shape orbital for the cube.

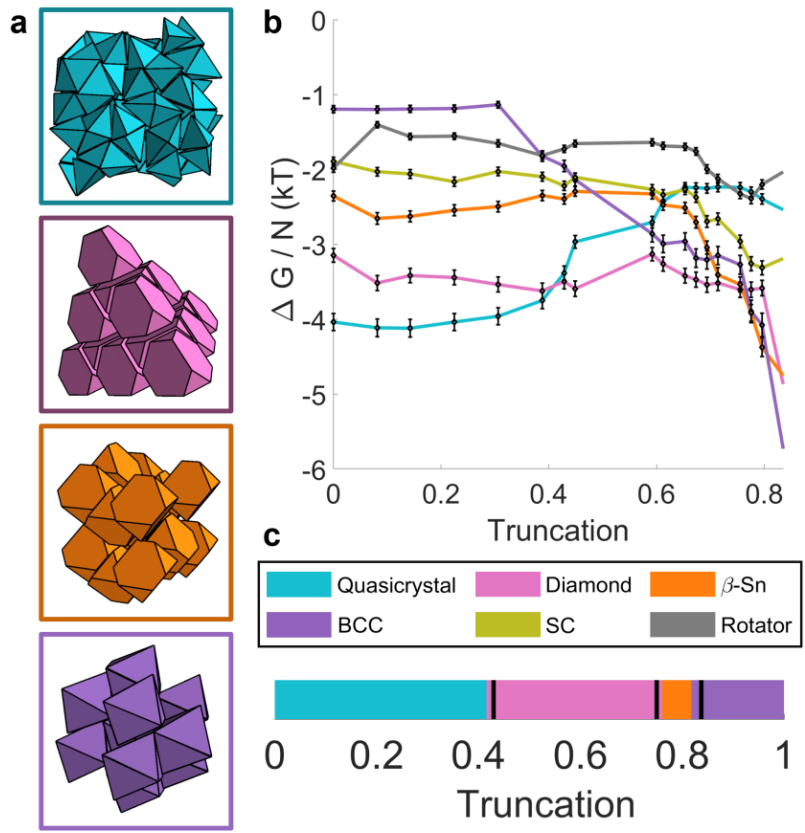

Figure 3: Entropic bond theory prediction for shape-driven solid-solid transition in a system of truncated tetrahedra. a). Snapshot of thermodynamically preferred structure with relative orientations between truncated tetrahedra predicted from theory. Colors correspond to the same lattice in (c). b) Lattice energy per particle used for phase diagram. We additionally computed the simple cubic (SC) and rotator crystal to show that other structures are not stable within this regime for the vertex-truncated tetrahedron shape family. c) Phase diagram for vertex-truncated tetrahedron shape family. Black lines indicate transitions between respective regions found in simulations (29). With increasing truncation, the stable phase transitions from quasicrystal approximant (QCA) to diamond to $\beta$-Sn to body-centered cubic (BCC). The LCSO approximation employs the first shape orbital for each polyhedron.

The entropic bond theory can also predict solid-solid transitions for a given particle shape as a function of packing fraction. We consider a system of regular trigonal bipyramids (TBP) for which a thermodynamic transition from a dodecagonal quasicrystal approximant (QCA) to a triclinic dimer lattice with increasing packing fraction was reported in Monte Carlo simulations (32). Representative snapshots of TBPs within the quasicrystal approximant and dimer lattice are shown in Fig. 4a-b, respectively. Our theory predicts the reported transition (Fig. 4c, black horizontal line in vertical color bar far right) at a packing fraction of $0.78 \pm 0.01$ as compared to the reported value of $0.79 \pm 0.008$ (32). Visualization of bonding orbitals for a TBP within the approximant and dimer phases indicates stark differences (Fig. 4d). To understand which bonding orbital pattern leads to the lower free energy above and below the transition, we examine the bonding free energy of a reference TBP and a neighboring TBP a distance $r$ away, at varying packing fractions. The bonding free energy curves reveal the appearance of longer-range entropic bonding *via* a second minimum at large $r$ at the transition from dimer to QCA lattice (Fig. 4c). This emergent phenomenon is consistent with calculations of the potential of mean force and torque (PMFT) from Monte Carlo simulations (22,23).

Complex atomic crystals have multiple Wyckoff positions, meaning that different positions within the crystal possess different local environments. The same is true for colloidal crystals, and entropic bond theory predicts this. As an example, we consider a system of hard pentagonal bipyramids (PBPs) that, in computer simulation, self-assembled from a fluid to a complex, multi-layered crystal with a 244-particle unit cell (oF244) upon crowding (11). Fig. 4e shows a snapshot of the PBPs within the oF244 lattice, colored according to their embedding layer L1, L2 or L3 within the crystal (Fig. 4f). As expected, the free energy of PBPs in the oF244 lattice is roughly similar across layers. Visualization of the bonding orbitals show strong pP localization on the PBP faces in L1 (Fig. 4g (left)), while pPs localize in a ring around the center of each PBP in L2 (Fig. 4g (center)) and are much less localized in L3 (Fig. 4g (right)). These results elucidate how diversity in local bonding environments can promote crystal complexity.

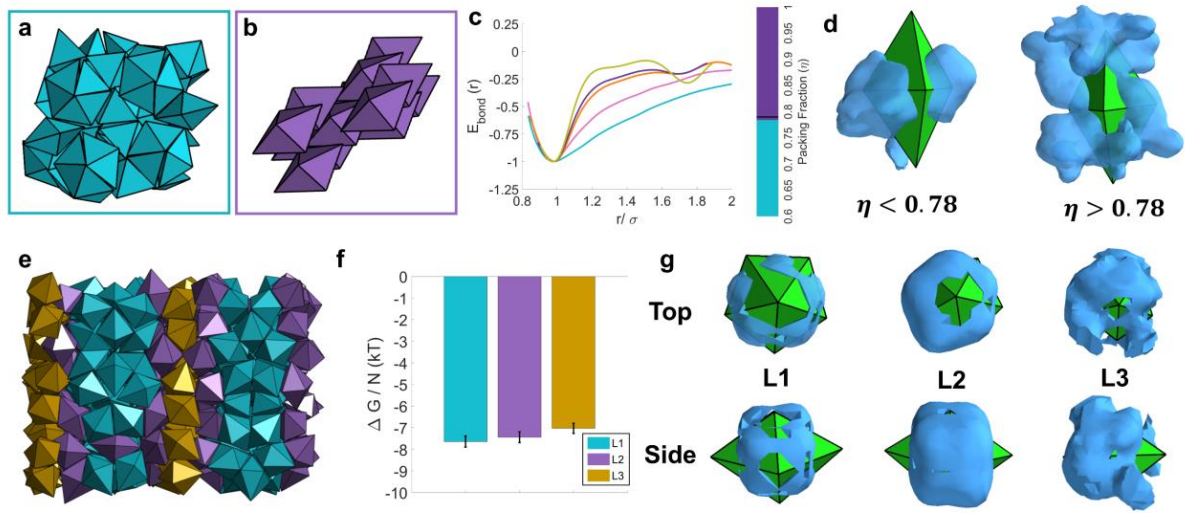

Figure 4: a) - d). Lattice prediction of packing-fraction driven solid-solid transition for trigonal bipyramids (TBP). Snapshot of a) 82-particle unit cell of quasicrystal approximant (QCA) at $\eta = 0.79$) and b) seven two-particle unit cells of the trigonal dimer packing at $\eta = 0.79$). c) Visualization of the entropic bonding orbitals around a reference TBP in QCA (left) and dimer (right) phases just below and above the solid-solid transition. The LCSO approximation employs the first two shape orbitals for every TBP. d) Calculated bonding (free) energy of a reference TBP and a neighboring TBP a distance $r$ away, at varying packing fractions $\eta$. At the previously reported (32) thermodynamic transition from the QCA to the dimer phase, $E_{bond}(r)$ undergoes a sharp transition from a single-well function for $\eta > 0.78$ to a double-well function for $\eta < 0.78$. The vertical color bar (far right) shows the lower free energy structure for TBP determined from lattice energy calculations (colors), and the reported transition from QCA to dimer phase from Monte Carlo simulation (black line) (32). e) - f) Local bonding environment prediction for a complex and multi-layered (oF244) lattice. e) Representative snapshot of oF244 lattice. f) Layer specific (L1, L2, L3) free energy per PBP in the oF244 lattice. g) While each PBP has similar free energy within each layer, visualization of the entropic bonding orbital at the same energy contour reveals that the local bonding environments differ greatly among L1 (left), L2 (center) and L3 (right). Top and bottom rows for g) show views from the top and side of the PBP, respectively. The LCSO calculation employs the first shape orbital for PBP.

**Discussion and Outlook**

We present a theoretical framework for predicting – *a priori* – thermodynamically preferred entropic colloidal crystal structures. We formulate a microscopic theory of entropy using a mathematical *ansatz* in which fictitious pseudoparticles are used to approximate excluded volume in a system of NPs. A self-consistent mean field derivation of a pP-NP interaction potential is combined with a convection-diffusion equation, resulting in an eigenvalue equation whose solution for a given colloidal crystal structure gives the entropy of the structure. By comparing solutions among many structures, we can select for the one with maximum entropy. We show that our framework predicts an effective entropic interaction between hard polyhedra that we quantify in terms of regions of high pP density and describe as an entropic bond. The entropic bond theory correctly predicts the thermodynamically preferred colloidal crystal structures for four different families of hard polyhedra as a function of shape or packing fraction. It also shows how different crystal structures – and different local environments within a crystal structure -- facilitate or hinder entropy maximization through visualization of bonding orbitals.

The entropic bond theory is different from, but complementary to, two conventional approaches for determining the assembly behavior of hard NPs – molecular simulations or Frenkel-Ladd free energy calculations (9,33–35). Molecular simulation (e.g. MC or MD) finds maximum entropy solutions (provided kinetics allows) by sampling millions of microstates in pursuit of the most probable state. Frenkel-Ladd calculations are used to compute free energies for given test crystals, but requires simulation: MC simulations are employed to compute an average free energy as a function of a given thermodynamic integration coupling parameter. The entropic bond theory requires no simulation of any kind. Rather, it employs an iterative, self-consistent solver akin to those used in electronic structure calculations to numerically solve an eigenvalue equation for the entropy of a crystal structure.

Entropic bond theory shares some similarities with chemical bonding theory, where significant theoretical developments have enabled calculation of ground-state energies by optimizing electron density throughout a crystal (36–39). In our theory, calculation of entropy in a colloidal crystal is performed by optimizing the density of a distribution of fictitious pseudoparticles, which permit quantification of excluded volume. pP density optimization is realized by maximizing the overlap of pP shape orbitals, which stabilizes the distance and orientation between neighboring anisotropic particles. By extension, solving for the equilibrium spatial distribution of pPs to find the most probable crystal structure of colloidal particles is akin to solving for the equilibrium spatial distribution of electrons to find the ground state crystal lattice of atoms. In other words, pPs act, mathematically speaking, as entropic analogs to electrons for entropically driven self-assembly. Whereas quantum mechanics is the language of chemical bonding, statistical mechanics is the language of entropic bonding. Table 1 provides an overview of these analogies.

| | Table 1: Chemical vs. Entropic Bonding | |
|---|---|---|
| | Chemical Bonding | Entropic Bonding |
| Core | Atomic Nucleus | Hard Shape |
| Mediator | Electrons | Pseudoparticles |
| Attraction | Nuclear-Electron | Shape-Pseudoparticle |
| Repulsion | Nuclear-Nuclear | Shape-Shape |
| | Electron-Electron | — |
| Orbitals | Atomic ($s,p,d,f$) | Shape |
| Bonding | Orbital Hybridization | Orbital Overlap |
| Governing Equation | Schrodinger | Equation 2 |

While predictive, the theory presented here suffers from the same limitations faced by electronic structure theory – namely, the need to guess lattices for comparison against each other. For ground state atomic crystals, where electronic structure theory is most useful, unit cells are comparatively small and simple. Only at non-zero temperatures do atomic crystal lattices become complex. For colloidal crystals, the same is true when comparing putative densest packings ("ground states" (40)) against the complex crystal structures observed away from infinite pressure. It is the latter we are most interested in, and guessing lattices may prove unwieldy. One particularly useful approach could be an adaptation of classical DFT for setting up gradient-descent type problems for optimization of particle positions rather than searching through a library of possible crystals, but further development is needed to determine its best uses (41).

Nevertheless, that there exists a theory based solely on entropy that, like electronic structure theory, is capable of predicting preferred crystal structures, is in itself important and profound. For one, it confirms that generalization of the term "bond" to describe the directional entropic forces (22) arising in crowded systems of hard polyhedra is mathematically appropriate and physically meaningful (23). Second, it helps explain why similar crystal structures can be observed on length scales orders of magnitude apart by demonstrating that, in the end, crystallization and self-assembly are agnostic to the origin of the forces between building blocks.

# References

1. M. A. Boles, M. Engel, D. V. Talapin, *Chem Rev* 116, 11220 (2016).

2. E. V. Shevchenko, D. V. Talapin, N. A. Kotov, S. O'Brien, C. B. Murray, *Nature* 439, 55 (2006).

3. A. M. Kalsin, *et al.*, *Science* 312, 420 (2006).

4. K. D. Hermanson, S. O. Lumsdon, J. P. Williams, E. W. Kaler, O. D. Velev, *Science* 294, 1082 (2001).

5. M. Grzelczak, J. Vermant, E. M. Furst, L. M. Liz-Marzan, *ACS Nano* 4, 3951 (2010).

6. A. K. Boal, *et al.*, *Nature* 404, 746 (2000).

7. P. F. Damasceno, M. Engel, S. C. Glotzer, *Science* 337, 453 (2012).

8. A. Haji-Akbari, *et al.*, *Nature* 462, 773 (2009).

9. A. Gantapara, J. De Graaf, R. Van Roij, M. Dijkstra, *Phys. Rev. Lett.* 11, 015501 (2013).

10. T. Schilling, S. Pronk, B. Mulder, D. Frenkel, *Phys. Rev. E* 71, 036138 (2005).

11. S. Lee, E. G. Teich, M. Engel, S. C. Glotzer, *P. Natl. Acad. Sci. USA* 116, 14843 (2019).

12. J. Henzie, M. Grunwald, A. Widmer-Cooper, P. L. Geissler, P. Yang, *Nat. Mater.* 11, 131 (2012).

13. Y. Nagaoka, *et al.*, *Nature* 561, 378 (2018).

14. T. C. Moore, J. A. Anderson, S. C. Glotzer, *Soft Mater.* (2021).

15. J. Zhu, *et al.*, *Nature* 387, 883 (1997).

16. Z. Cheng, W. B. Russel, P. M. Chaikin, *Nature* 401, 893 (1999).

17. L. Onsager, *Ann. NY Acad. Sci* 51, 627 (1949).

18. W. W. Wood, J. D. Jacobson, *J. Chem. Phys* 27, 1743956 (1957).

19. B. J. Alder, T. E. Wainwright, *J. Chem. Phys* 27, 1743957 (1957).

20. D. Frenkel, *Physica A* 263, 26 (1999).

21. D. Frenkel, *Nat. Mater.* 14, 9 (2015).

22. G. van Anders, D. Klotsa, N. K. Ahmed, M. Engel, S. C. Glotzer, *P. Natl. Acad. Sci. USA* 111, E4812 (2014).

23. E. S. Harper, G. van Anders, S. C. Glotzer, *P. Natl. Acad. Sci. USA* 116, 16703 (2019).

24. W. Kohn, L. J. Sham, *Phys. Rev* 140, A1133 (1965).

25. I. N. Levine, *Quantum Chemistry* (Pearson, 2014), 7th edn.

26. Z. Zhang, S. C. Glotzer, *Nano Lett* 4, 1407 (2004).

27. S. C. Glotzer, M. Solomon, *Nat. Mater.* 6, 557 (2007).

28. M. A. Carignano, I. Szleifer, *J. Chem. Phys* 98, 5006 (1998).

29. P. F. Damasceno, M. Engel, S. C. Glotzer, *ACS Nano* 6, 609 (2011).

30. M. Nic, J. Jirat, B. Kosata, *Iupac. compendium of chemical terminology* (the "gold book"), second edn.

31. F. Smallenburg, L. Filion, M. Marechal, M. Dijkstra, *P Natl Acad Sci USA* 109, 17886 (2012).

32. A. Haji-Akbari, M. Engel, S. C. Glotzer, *Phys. Rev. Lett* 107, 215702 (2011).

33. U. Agarwal, F. A. Escobeda, *Nat. Mater.* 10, 230 (2011).

34. A. Haji-Akbari, M. Engel, S. C. Glotzer, *J. Chem. Phys* 135, 194101 (2011).

35. C. Vega, E. G. Noya, *J. Chem. Phys* 127, 154113 (2007).

36. J. Yang, *et al.*, *Science* 345, 640 (2014).

37. H. Park, A. Millis, C. A. Marianetti, *Phys. Rev. B* 90, 235103 (2014).

38. Y. Cao, *et al.*, *Nature* 556, 43 (2018).

39. X. Gonze, *et al.*, *Computer Physics Communications* 205, 106 (2016).

40. J. de Graaf, L. Filion, M. Marechal, R. van Roij, M. Dijkstra, *J. Chem. Phys* 137, 214101 (2012).

41. J. Wu, *Molecular Modeling and Simulation (Applications and Perspectives)* (Springer, 2017).

## Acknowledgments

This work was supported in part by a grant from the Simons Foundation (256297, SCG) and by the Department of the Navy, Office of Naval Research under ONR award number N00014-18-12497. Additionally, this research was supported by the Office of the Undersecretary of Defense for Research and Engineering (OUSD(R&E)), Newton Award for Transformative Ideas during the COVID-19 Pandemic, Award number HQ00342010030. This work used the Extreme Science and Engineering Discovery Environment (XSEDE), which is supported by National Science Foundation Grant ACI-1053575; XSEDE Award DMR 140129. Computational resources and services were also provided by Advanced Research Computing (ARC-TS) at the University of Michigan, Ann Arbor, MI. T.V acknowledges Vyas Ramasubramani for helpful discussions regarding the clarity of both manuscript and figures.